\documentclass[12pt]{article}
\usepackage{epsfig}
\usepackage{graphicx}

\def\beq{\begin{equation}}
\def\eeq#1{\label{#1}\end{equation}}
\def\eeqn{\end{equation}}

\def\beqa{\begin{eqnarray}}
\def\eeqa#1{\label{#1}\end{eqnarray}}
\def\eeqan{\end{eqnarray}}

\let\bar=\overbar

\def\Dslash{\not{\hbox{\kern-4pt $D$}}}
\def\dslash{\not{\hbox{\kern-2pt $\del$}}}

\def\msb{{\bar{\ssstyle M \kern -1pt S}}}

\usepackage{fancyhdr,graphicx}
\def\Title#1{\begin{center} {\Large {\bf #1} } \end{center}}

\begin{document}

\Title{Inferring neutron-star properties from gravitational-wave signals of binary mergers}

\bigskip\bigskip


\begin{raggedright}

{\it 
Andreas Bauswein$^{1}$~~Nikolaos Stergioulas$^{1}$~~and Hans-Thomas Janka$^{2}$\\
\bigskip
$^{1}$
Department of Physics,
Aristotle University of Thessaloniki,
GR-54124 Thessaloniki,
Greece\\
\bigskip
$^{2}$
Max-Planck-Institut f\"ur Astrophysik, 
D-85748 Garching,
Germany\\
}

\end{raggedright}

\section{Introduction}
The relativistic stellar structure equations uniquely link the equation of state of high-density matter to the mass-radius relation of non-rotating neutron stars. This implies that a measurement of the mass-radius relation or a sufficient number of individual points on the mass-radius relation determine the high-density equation of state. The upcoming gravitational-wave detectors are designed to be sensitive for gravitational radiation with frequencies in the range from several 10 Hz to a few kHz. Merging neutron stars are among the prime targets of the upcoming gravitational-wave detectors, since the frequencies of the orbital motion and the postmerger oscillations fall in mentioned frequency range (see e.g.~\cite{2012LRR....15....8F} for a review). Neutron stars in binary systems are doomed to merge since the orbital motion generates gravitational waves, which extract energy and angular momentum from the system and result in an inspiralling motion of the binary components. Detection rates between 0.4 and 
400 events per year are estimated for the upcoming instruments~\cite{2010CQGra..27q3001A}. For near-by events with sufficient signal-to-noise ratio there are two complementary ways of inferring stellar properties from gravitational-wave detections. During the last cycles of the inspiral phase finite-size effects are imprinted on the gravitational-wave signal and may be extracted if the exact shape of the signal is understood (see e.g.~\cite{2013arXiv1306.4065R}). Considering merger simulations with a large sample of candidate equations of state, the outcome of a merger for the most likely binary mass range is the formation of a differentially rotating neutron-star merger remnant~\cite{2012PhRvL.108a1101B,2012PhRvD..86f3001B,2013PhRvL.111m1101B}. Since the structure of this remnant is highly sensitive to the equation state, the remnant's oscillations~\cite{2011MNRAS.418..427S} are very characteristic for the equation of state~\cite{2012PhRvL.108a1101B,2012PhRvD..86f3001B}. In this contribution we describe how 
the dominant oscillation frequency of the remnant can be employed to infer the radii of non-rotating neutron stars. We also mention the expected accuracy of binary mass determinations from the inspiral gravitational-wave signal, which are required to interpret the postmerger signal.

\section{Radius measurements}
The dominant oscillation frequency of the merger remnant occurs as a pronounced peak in the kHz range of the gravitational-wave spectrum. This peak is a robust feature in all models which lead to the formation of a neutron-star merger remnant and which do not promptly collapse to a black hole. Simulations with a fixed total binary mass, e.g. $M_\mathrm{tot}=2.7~M_\odot$, reveal that the frequency $f_{\mathrm{peak}}$ of the postmerger peak correlates very well with the radii $R_{\mathrm{NS}}$ of non-rotating neutron stars of a chosen mass $M_{NS}$~\cite{2012PhRvL.108a1101B,2012PhRvD..86f3001B}. (The radius of a non-rotating neutron star for a given equation of state is a convenient way of characterizing the equation of state and thus the gravitational-wave emission of a model employing this equation of state.) In principle, any mass $M_{NS}$ can be chosen to relate $R(M_{NS})$ with $f_{\mathrm{peak}}$. Empirically, it turns out that the relation between $f_{\mathrm{peak}}$ and $R_{\mathrm{NS}}$ becomes 
particularly tight if $M_{NS}$ is somewhat larger than the mass of the inspiralling neutron stars, i.e. larger than $M_\mathrm{tot}/2$ (here we primarily focus on equal-mass binaries). This does not come as a surprise since the maximum density in the remnant 
exceeds the one of the initial neutron stars and is more comparable to the densities in a non-rotating neutron star with $M_{NS} \approx 1.6~M_\odot$. Therefore, the radius of a nonrotating NS with $1.6~M_\odot$ is particularly suitable to characterize the stellar structure of the massive rotating merger remnant and its oscillation frequencies. A more natural choice would be $M_{NS}=M_\mathrm{tot}/2$, which means characterizing the remnant by the radius of the inspiralling stars. Also in this case a relation between $R(M_{NS})$ and $f_{\mathrm{peak}}$ is found, but with a somewhat larger scatter compared to the use of a larger fiducial neutron-star mass $M_{NS}$~\cite{2012PhRvD..86f3001B}.

\begin{figure}[htb]
\centering
\includegraphics[width=0.6\textwidth]{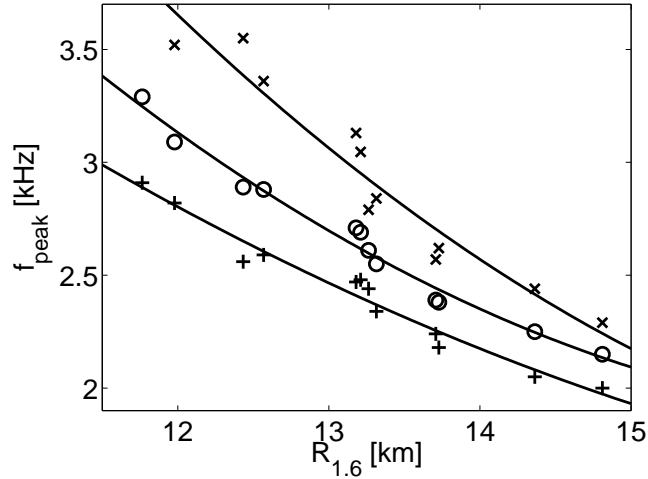}
\caption{Gravitational-wave frequency of the dominant oscillation of the postmerger remnant for different equations of state as function of the radius of nonrotating neutron stars with 1.6~$M_\odot$. Shown are simulation results for total binary masses of 2.4~$M_\odot$ (plus signs), 2.7~$M_\odot$ (circles) and 3.0~$M_\odot$ (crosses). Solid lines are least-square fits to the empirical relations for fixed total binary masses.}
\label{fig:fpeak}
\end{figure}
Our procedure of inferring NS radii from NS binary mergers relies on the inversion of the relation between the fiduical NS radius $R_\mathrm{NS}$ and the dominant GW frequency $f_{\mathrm{peak}}$ of the postmerger phase. As already pointed out in~\cite{2012PhRvL.108a1101B,2012PhRvD..86f3001B}, there exist different relations for different total binary masses $M_\mathrm{tot}$. This is illustrated in Fig.~\ref{fig:fpeak}, which displays simulation results for different temperature-dependent high-density equations of state. The temperature dependence is an important feature of these equations of state for the quantitative results of the merger simulations~\cite{2010PhRvD..82h4043B}; details on these equations of state can be found in~\cite{2014PhRvD..90b3002B}. Binaries with mass ratios different from unity yield to a good approximation the same $f_{\mathrm{peak}}$ as the equal-mass binary of the same total mass. Therefore, a corresponding  $R_\mathrm{NS}$-$f_{\mathrm{peak}}$ relation with a fixed mass ratio 
$q$ unequal unity can be very well approximated by the relation given by the symmetric binary with the same total mass~\cite{2012PhRvL.108a1101B,2012PhRvD..86f3001B}. Our recipe for determining NS radii thus only relies on the capability of GW detectors to reveal the total binary mass from the inspiral phase and of determining the dominant frequency of the postmerger phase. The mass measurement will allow to choose the corresponding $f_\mathrm{peak}-R_\mathrm{NS}$ relation depending on $M_{\mathrm{tot}}$. Basically, this relation could be computed after $M_\mathrm{tot}$ is known from a measurement by performing a number of additional simulations for the different EoSs with the measured total binary mass. Alternatively, the relation for a given $M_\mathrm{tot}$ could be obtained by a simple interpolation between two given $f_\mathrm{peak}-R_\mathrm{NS}$ relations for two different total binary masses.

The capabilities of the upcoming GW detector network to measure the binary parameters of NS mergers have been explored for instance in~\cite{2014ApJ...784..119R}. These simulations have shown that for a network signal-to-noise ratio (SNR) of 20, the total binary mass is determined from the GW inspiral phase with an accuracy of about 1.5 per cent at the 95 per cent confidence level. As demonstrated in~\cite{2014PhRvD..90f2004C} the dominant postmerger oscillation frequency can be measured only for merger events within about 25~Mpc. For such a detection, the SNR of the inspiral phase is expected to be five to ten times higher. Anticipating that the precision of the $M_\mathrm{tot}$ determination scales linearly with 1/SNR, the total binary mass will be known practically exactly for the purpose of choosing the corresponging $f_\mathrm{peak}-R_\mathrm{NS}$ relation. 

While small asymmetries of the binary system do not change the $f_\mathrm{peak}-R_\mathrm{NS}$ relation for a given $M_\mathrm{tot}$, one may want to check that the mass ratio of a detected merger event does not deviate too strongly from unity. For SNR=20 the individual masses of the binary system are recovered within $\sim$10 per cent with an even higher precision for larger asymmetries~\cite{2014ApJ...784..119R}. In the case of a near-by merger event allowing a  measurement of $f_\mathrm{peak}$, the individual component masses will thus be determined within a few per cent. This implies that even $f_\mathrm{peak}-R_\mathrm{NS}$ relations for given $(M_\mathrm{tot},q)$ combinations can be employed to measure NS radii, which may prove useful especially for systems which strongly deviate from $q=1$. In summary, these considerations show that the binary masses of near-by NS mergers will be measured sufficiently accurately from the inspiral phase and will only introduce negligble errors to the radius determination.

The main source of error in a neutron-star radius measurement from the postmerger gravitational-wave signal is given by the scatter in the empirical $f_\mathrm{peak}-R_\mathrm{NS}$ relation. It is not possible to predict in which way the $f_\mathrm{peak}$ of the true equation of state deviates from the empirical $f_\mathrm{peak}-R_\mathrm{NS}$ relation. Therefore, one has to adopt a possible error given by the largest deviation of any theoretically possible model from the empirical relation. Here we implicitly assume that the sample of candidate equations of state contains extreme models, which yield deviations from the fit which are larger than the one of the true equation of state. It can be seen in Fig.~\ref{fig:fpeak} that the error of the radius measurement is of the order of about 200 meters for $M_\mathrm{tot}=2.7~M_\odot$. For larger or smaller total binary masses a determination of $R(1.6~M_\odot)$ yields a somewhat larger error, which is understandable considering the discussion above about the optimal choice of $M_{NS}$. Indeed, a smaller scatter is found for the $M_\mathrm{tot}=2.4~M_\odot$ data if $f_\mathrm{peak}$ is described as a function of radii of $\sim 1.35~M_\odot$ neutron stars~\cite{2012PhRvD..86f3001B}. Similarly, choosing  $R(1.8~M_\odot)$ to describe the $M_\mathrm{tot}=3.0~M_\odot$ data leads to a tighter relation than the one shown in Fig.~\ref{fig:fpeak}. Note that comparisons to the simulations of other groups confirm the quantitative results for the peak frequencies, e.g.~\cite{2005PhRvD..71h4021S,2013PhRvD..88d4026H,2014PhRvL.113i1104T,2015PhRvD..91f4027K,2015PhRvD..91f4001T} (see also discussion in~\cite{2012PhRvD..86f3001B}).

The detectability of the dominant oscillation frequency was discussed in~\cite{2014PhRvD..90f2004C}. Therein it was shown that the frequency determination in a gravitational-wave detection contributes only a very small error to the radius measurement.

Finally, we note that a measurement of $f_\mathrm{peak}$ allows a rough radius determination even without the precise knowledge of the total binary mass since the fits for different binary masses do not differ too much (see Fig.~\ref{fig:fpeak}). With reasonable assumptions about the possible $M_\mathrm{tot}$ range, e.g. from the precisely measured chirp mass or from statistical arguments derived from pulsar observations, NS radii may be determined within $\sim$km precision.

\section{Summary and outlook}
In this contribution we presented the idea to infer the radii of non-rotating neutron stars from the detection of the dominant oscillation frequency of the gravitational-wave signal of neutron-star merger remnants. We shed light on the theoretical background of the underlying relations and discussed several sources of error. We point out that the detection of the postmerger gravitational-wave emission of two events with different total binary masses may be employed to estimate the emission properties close to the threshold binary mass leading to the gravitational collapse of the remnant. The properties at the threshold in turn allow the determination of the maximum mass of non-rotating neutron stars and the radius of the maximum-mass configuration~\cite{2014PhRvD..90b3002B}. Also the direct observation of a prompt black-hole formation after a merger and the measured total binary mass may constrain the maximum mass~\cite{2013PhRvL.111m1101B}.

\subsection*{Acknowledgement}

A.B. expresses his thanks to the organizers of the CSQCD IV conference for providing an 
excellent atmosphere which was the basis for inspiring discussions with all participants.
He has greatly benefitted from this. A.B. is a Marie Curie Intra-European Fellow within the 7th European Community Framework Programme (IEF 331873) and is grateful for travel support from the HadronPhysics3 Programme. Computer time at the Rechenzentrum Garching of the Max-Planck-Gesellschaft, the Max Planck Institute for Astrophysics, and the Cyprus Institute under the LinkSCEEM/Cy-Tera project is acknowledged.


\end{document}